# Real-time Hall-effect detection of current-induced magnetization dynamics in ferrimagnets

G. Sala[1*], V. Krizakova[1], E. Grimaldi[1], C.-H. Lambert[1], T. Devolder[2], and P. Gambardella[1*]

[1]Department of Materials, ETH Zurich, 8093 Zurich, Switzerland

[2]Centre de Nanosciences et de Nanotechnologies, CNRS, Université Paris-Sud, Université Paris-Saclay, 91405 Orsay Cedex, France

*email : giacomo.sala@mat.ethz.ch; pietro.gambardella@mat.ethz.ch

**ABSTRACT**

Measurements of the transverse Hall resistance are widely used to investigate electron transport, magnetization phenomena, and topological quantum states. Owing to the difficulty of probing transient changes of the transverse resistance, the vast majority of Hall effect experiments are carried out in stationary conditions using either dc or ac currents. Here we present an approach to perform time-resolved measurements of the transient Hall resistance during current-pulse injection with sub-nanosecond temporal resolution. We apply this technique to investigate in real-time the magnetization reversal caused by spin-orbit torques in ferrimagnetic GdFeCo dots. Single-shot Hall effect measurements show that the current-induced switching of GdFeCo is widely distributed in time and characterized by significant activation delays, which limit the total switching speed despite the high domain-wall velocity typical of ferrimagnets. Our method applies to a broad range of current-induced phenomena and can be combined with non-electrical excitations to perform pump-probe Hall effect measurements.

**INTRODUCTION**

The broad family of Hall effects includes phenomena of ordinary, anomalous[1], planar[2,3], topological[4,5], and quantum[6–8] origin. These effects have become standard tools for benchmarking the physics of metallic, semiconducting, and topological materials as well as the functionality of electronic and spintronic devices. The anomalous Hall effect (AHE), for example, allows for probing the emergence of magnetically-ordered phases[1,9–11], field-[12] and current-induced magnetization reversal[13–15], domain wall motion[16], and spin-orbit torques (SOTs) [17–19]. Measurements of the transverse resistance also provide insight into magnetoresistive phenomena, such as the planar Hall effect and spin Hall magnetoresistance, which can be used to track the response of antiferromagnets and magnetic insulators to applied magnetic fields, currents, and heat[20–22]. Extending these measurements to the time domain would enable access to the dynamics of a vast range of electronic and magnetic systems. As is well-known, the ordinary and planar Hall effects are widely employed in sensors for the detection of magnetic fields and microbeads[23–25], and have a frequency bandwidth extending up to several GHz (Refs. [26,27]). However, there are only few examples of time-resolved ($\mu$s-ns) measurements of the magnetization dynamics using the Hall effect, which are limited to observations of laser-induced heating[28] and the transit of domain walls[29,30].

Here, we present an all-electrical technique suitable for systematic real-time measurements of any kind of transverse magnetoresistance in devices with current flowing in-plane. The key idea consists in disentangling the tiny magnetic Hall signal from the large non-magnetic background by minimizing the current leakage in the sensing arms of the Hall cross. This approach, which relies on the counter-propagation of electric pulses, is well adapted for radio-frequencies and proves particularly useful for fast excitations, e.g., ns- and sub-ns-long pulses. We demonstrate the capability of this technique by studying the magnetization dynamics triggered by SOTs[17] in ferrimagnetic GdFeCo dots patterned over a Pt Hall bar. In our detection scheme, the ns-long pulses do not only generate the perturbation on the magnetization but also serve as the tool for tracking the magnetic response, including single-shot switching events. This capability opens up the possibility of performing systematic time-resolved Hall measurements of current-induced excitations in a broad variety of planar devices and provides access to stochastic events.

Ferrimagnets have recently attracted considerable attention due to the enhanced SOT efficiency[31–33] and the extraordinary high current-induced domain-wall velocity[34–36] attained, respectively, at the magnetization and angular-momentum compensation points. These properties make them promising candidates for the realization of fast and energy-efficient spintronic devices[36]. However, the current-driven magnetization dynamics in these systems has been investigated only using magneto-optical pump-probe methods[36,37], which do not provide information on stochastic events. Our time-resolved AHE measurements show that the reversal of the magnetization in GdFeCo evolves in different phases, which comprise an initial quiescent state, the fast reversal of the magnetization, and the subsequent settling in the new equilibrium state without ringing effects. Despite the high domain-wall velocity attained by ferrimagnets, we find that the total switching time is severely affected by an initial activation phase, during which the magnetization remains quiescent. We associate this phase, which has not been reported so far in ferrimagnets, with the time required to nucleate a reversed domain assisted by Joule heating. The single-shot AHE traces reveal the existence of broad distributions of the nucleation and reversal times and disclose the stochastic character of the SOT-induced dynamics, which is not accessible to pump-probe techniques. Our measurements further show that the domain nucleation time can be substantially reduced by increasing the current amplitude, leading to a minimum of the critical switching energy for pulses of reduced length.

**RESULTS**

**Time-resolved anomalous-Hall-effect measurements.**

Electrical time-resolved measurements using the Hall effect, or any form of transverse magneto-resistance, suffer from the difficulty of generating a detectable Hall signal without spoiling the signal-to-noise ratio. The main obstacle is the current shunting into the sensing line of the Hall cross, caused by the finite electric potential at its center. When a pulse reaches the cross, a portion of the current flows through the transverse arms (along $\pm y$ in the top panel of Fig. 1a), thus producing a spurious electric potential associated with the resistance of the leads. This potential is much larger than the signal of magnetic origin and hinders its detection. A limitation remains even in differential measurements because the unavoidable asymmetry of the leads introduces a finite differential offset[23] that can saturate the dynamic range of the Hall voltage amplification stage. These problems do not exist in standard dc measurements as the current leakage is countered by the high input impedance of the measuring instrument. At high frequency, however, impedance matching requires a low resistance (50 Ohm) at the input port of the instrument, usually an oscilloscope.

The approach that we introduce here consists in injecting two counter-propagating rf pulses with amplitude $\left|\frac{V_P}{2}\right|$ and opposite polarity, as depicted in the bottom panel of Fig. 1a. Provided that these pulses reach the center of the cross at the same time and have the same amplitude, a virtual ground is forced there. The virtual ground limits the spread of the current because the voltage drop on the entire sensing line (Hall arm, cable, and input impedance of the oscilloscope) is ideally zero. The synchrony of the two balanced pulses, generated by a balun power divider, is ensured by the symmetry of the paths connecting the balun to the device, as schematized in Fig. 1b. Thanks to the opposite polarity of the pulses, the current flows along the $x$ direction, with double magnitude relative to the current produced by a single pulse of amplitude $\frac{V_P}{2}$, and sign determined by the polarity of the pulses. The current generates time-dependent transverse Hall voltages, $V_+$ and $V_-$, which are pre-amplified and acquired by a sampling oscilloscope triggered by an attenuated portion of the original pulse. If no change of the magnetization occurs during the pulses, the magnetic signal mimics the shape of the pulse. A deviation from this reference signal is the signature of ongoing magnetization dynamics. In the specific case discussed below, the transverse voltage stems from the AHE and its change over time gives access to the out-of-plane component of the magnetization. We note that, in the more general situation of asymmetric Hall crosses, our technique allows for compensating detrimental resistance offsets by tuning the relative amplitude of the counter propagating pulses. This capability is unique to our approach and cannot be implemented in time-resolved differential Hall measurements[30]. We also remark that the main additional component to the setup required by our approach is the balun divider, which is a simple and affordable circuit element. More details about the electric circuit, including the rf and dc sub-networks, sensitivity,

resistance offsets compensation, and time-resolution are discussed in the Methods and in Supplementary Notes 1, 2, and 5.

**Switching dynamics of ferrimagnetic dots.**

We adapted this concept to investigate the SOT-induced magnetization switching of 15-nm-thick, 1-μm-wide $Gd_{30}Fe_{63}Co_7$ dots with perpendicular magnetization, patterned on top of a 5-nm-thick Pt Hall bar (see Fig. 1c,d, Methods, and Supplementary Note 3). The compensation temperature of the ferrimagnetic dots is below room temperature, such that the net magnetization and AHE are dominated by the magnetic moments of Fe and Co. Therefore, in our room-temperature measurements the current-induced switching in the presence of an in-plane static magnetic field has the same polarity as in perpendicularly-magnetized ferromagnets with a Pt underlayer[14,17]. Specifically, the parallel alignment of current and field favours the down state of the magnetization, whereas the antiparallel orientation promotes the up state, which correspond to negative and positive anomalous Hall resistance, respectively.

The differential signal $V_+ - V_-$ is determined by the magnetization orientation, which changes with time during a switching event. Figure 2a shows the switching traces obtained by measuring $V_+ - V_-$ during the reversal of a GdFeCo dot for different pulse amplitudes. In order to minimize spurious contributions to the magnetic signal, a background signal was recorded by fixing the magnetization in the initial state, either "up" or "down", and subtracted from the data. The down-up and up-down switching traces obtained by averaging over 1000 pulses are shown as red and blue lines, respectively. The black lines represent a reference trace obtained by subtracting two background measurements corresponding to the magnetization pointing up and down. This reference trace describes the maximum excursion of the Hall voltage during a current pulse (see Supplementary Note 4 for more details). The deviation of the switching traces from the top and bottom reference levels corresponds to the change of the out-of-plane magnetization driven by the SOTs during the 20-ns-long current pulse. Dividing the switching traces by the corresponding reference trace provides the normalized magnetic time traces shown in Fig. 2b-e. In these average measurements, the transition between the top and bottom reference levels of the switching trace is sufficiently clear such that the normalization by the reference trace is not strictly required. The latter, however, is important to highlight the switching in single-shot measurements, which will be presented later on.

The measurements in Fig. 2b-e allow us to electrically probe the time-resolved SOT-induced dynamics in planar devices, which so far has been achieved only by X-ray and magneto-optical techniques[36–39]. We find that the switching dynamics of the ferrimagnetic dots comprises three phases: an initial quiescent state, the reversal phase, and the final equilibrium state, with the magnetization remaining constant both before and after the reversal. Both the quiescent and reversal phase present stochastic components. The observation of a long quiescent phase challenges the common assumption that the magnetization reacts instantaneously to the SOT owing to the orthogonality between the initial magnetization direction and the torque[40–42], unlike the spin-transfer torque between two collinear magnetic layers[43]. Instead, our measurements show that the duration of this phase can be comparable to the pulse length. The quiescent phase is a characteristic of the thermally-activated regime, in which thermal fluctuations assist the switching and lead to a stochastic delay time. Because of the relatively high perpendicular anisotropy of the ferrimagnetic dots (see Supplementary Note 3), the thermal activation plays a role up to current density of the order of $1.5 \times 10^{12}$ A m$^{-2}$, similar to the switching of high-coercivity ferromagnetic nanopillars by spin transfer torque[44]. By increasing the pulse amplitude or the in-plane field, the duration of quiescent phase is significantly reduced as the switching dynamics approaches the intrinsic regime (see Fig. 2b-e and the following sections).

**Single-shot measurements**

Although the averaging process improves the quality of the traces, it conceals the stochastic nature of the dynamics. Here, we show that our technique provides sufficient signal-to-noise contrast to detect individual reversal events in Hall devices. By using the procedure outlined above, we measured single-shot switching traces for different in-plane magnetic fields and pulse amplitudes, as shown in Fig. 3 for three representative voltages. The single-shot traces are qualitatively similar to the average traces. However, the duration of the quiescent and transition phases varies significantly from trace to trace. By fitting each trace to a piecewise linear function, we define $t_0$ as the duration of the initial quiescent phase during which the normalized Hall voltage remains close to 1 (0) before the up-down (down-up) reversal (see Methods). In the following, we refer to $t_0$ as the nucleation time, arguing that the quiescent phase is associated with the reversal of a seed domain[38,45,46], in analogy to measurements performed on ferromagnetic tunnel junctions[47]. Additionally, we designate the duration of the transition between the up-down or down-up magnetization levels as the transition time $\Delta t$ (Ref. [48]). The total switching time is thus given by $t_0 + \Delta t$.

To gain insight into the stochastic variations of $t_0$ and $\Delta t$, we recorded a set of 1000 individual traces for several values of the applied in-plane field *B* and voltage *V*. Figure 4 shows the statistical distributions of $t_0$ and $\Delta t$ obtained at representative fields and pulse amplitudes. The comparison between the single-shot statistics in Fig. 4 and the averaged traces in Fig. 2 reveals that the duration of the quiescent phase is systematically underestimated in the average measurements relative to the mean $\bar{t}_0$, whereas the duration of the transition phase is systematically overestimated relative to the mean $\overline{\Delta t}$. The deviation of the times deduced from the average measurements relative to $\bar{t}_0$ and $\overline{\Delta t}$ can reach up to -25% and 60%, respectively. The quantitative disagreement is determined by the superposition of widely-distributed nucleation events. As shown by the average curves at the bottom of Fig. 3, the large spread of the nucleation events anticipates the starting point of the average dynamics and, at the same time, broadens the apparent switching duration. Therefore, only single-shot measurements can accurately quantify the full switching dynamics, including the variability of events as well as the duration of the nucleation and transition phases, and their distributions.

The data reported in Fig. 4 show that $t_0$ approximately follows a normal distribution, as expected from random events. In contrast, $\Delta t$ has a significant positive skew with the mean $\overline{\Delta t}$ shifted towards the shorter times. Moreover, $\bar{t}_0$ and its standard deviation decrease strongly upon increasing either the pulse amplitude or the field, whereas $\overline{\Delta t}$ shows only a moderate dependence on the voltage. These distinct statistical distributions and dependencies are the signature of different physical processes underlying the initial phase and the transition phase of the reversal. Doubling the pulse amplitude or field leads to a ~10-fold reduction of $\bar{t}_0$, consistently with an activated domain nucleation process that is promoted by SOTs and assisted by the in-plane field[47] and thermal fluctuations.

In contrast with $\bar{t}_0$, the effect of the in-plane field on $\overline{\Delta t}$ is negligible. This observation supports the interpretation of $\Delta t$ in terms of domain-wall depinning and propagation time, since, for the fields used in this study, the domain wall mobility is saturated at the maximum value expected for Néel walls[49,50]. On the other hand, stronger pulses are expected to ease the depinning of domain walls and increase their speed, in accordance with the reduction of $\overline{\Delta t}$ at larger voltages. Consistent with our analysis, $\Delta t$ can be interpreted as the time required for the seed domain to expand across the entire area of the dot. Therefore, the inverse of $\Delta t$ provides an upper limit to the domain wall velocity in our devices. The average domain wall velocity estimated from the mean of the distributions reaches several hundreds of m/s, whereas the peak velocity can be as large as 4 km/s. Such a high speed is in line with the velocities estimated by measuring the domain wall displacements in GdFeCo following the injection of current pulses[34–36]. Further improvements of the domain wall velocities have been demonstrated by tuning the stoichiometry and transient temperature of GdFeCo so as to approach the angular momentum compensation point[35]. Our measurements demonstrate that the nucleation phase, characterized by a long delay time $t_0$, is the real bottleneck of the SOT-induced switching dynamics of ferrimagnets. Therefore, the efficient operation of ferrimagnetic devices based on SOTs requires strategies to reduce the initial quiescent phase and mitigate the associated stochastic effects.

**Intrinsic and thermally activated switching regimes.**

Measurements of the threshold switching voltage $V_c$ as a function of the pulse duration $t_P$ evidence the existence of two switching regimes[40], as shown in Fig. 5 (see also Supplementary Note 6). Above approximately 5 ns, $V_c$ changes weakly with $t_P$, which is a signature of the thermally-assisted reversal[40,44] and reveals the importance of thermal effects for the typical pulse lengths and amplitudes used in this study ($t_P = 20$ ns). On the other hand, the critical voltage increases abruptly for $t_P \lesssim 3$ ns, as expected in the intrinsic regime where the switching speed depends on the rate of angular momentum transfer from the current to the magnetic layer. Indeed, in this regime, $V_c$ scales proportionally to $1/t_P$ (see Supplementary Fig. S7). Switching with $t = 300$ ps (equivalent average domain wall speed > 3.3 km/s, under the assumption $t_0 \approx 0$) demonstrates that the quiescent phase can be suppressed by strongly driving the magnetization. In this case, the SOTs alone are sufficiently strong to drag the magnetization away from the equilibrium position and induce the nucleation of a domain against the energy barrier without substantial thermal aid. Finite element simulations support this point by showing that the temperature rise times in our devices are larger than 2 ns.

Importantly, the suppression of the quiescent phase requires more intense pulses but does not imply a larger energy consumption because the threshold energy density decreases by more than 4 times upon reducing $t_P$ from 20 ns to < 1 ns (see Fig. 5). This favorable trend highlights the advantage of using materials for which the fast dynamics does not require excessively large current densities. We note that the current densities used in this study are compatible with previous results obtained on GdCo (Ref. [36]). In that work the current density at 300 ps is approximately $1.05 \times 10^{12}$ A m$^{-2}$, whereas in our devices with three times larger GdFeCo thickness the threshold current density reaches $3.6 \times 10^{12}$ A m$^{-2}$. For 20-ns-long pulses, this value reduces to $0.82 \times 10^{12}$ A m$^{-2}$. On the other hand, a more stringent comparison of our findings with the measurements reported in Ref. [36] is not straightforward because the device geometries, the materials and their magnetic properties are dissimilar.

**Sensitivity and temporal resolution.**

Finally, we present considerations on the sensitivity and time resolution of our technique that apply to all conductors with a finite transverse resistivity $\rho_{xy}$. In all generality, we assume that $\rho_{xy} \neq 0$ only in a finite region of the Hall cross (the "magnetic dot"). The Hall voltage generated by two counter-propagating voltage pulses of opposite amplitude $V_P/2$ and $-V_P/2$ is given by $V_+ - V_- = f \frac{\rho_{xy}}{t} \frac{V_P}{R_I}$, where $t$ is the thickness of the dot, $R_I$ the resistance of the injection line, and $f$ a sensitivity factor ($< 1$) that depends on the ratio between the area of the dot and the Hall cross as well as on the inhomogeneous current distribution within the device. An equivalent circuit model of the Hall cross and sensing apparatus shows that the differential Hall signal $S$ measured at the input ports of the oscilloscope is the result of the amplified voltage partition between the two branches of the sensing line, each having a resistance $R_S$, and the input resistance of the amplifier $R_A$:

$$S = 2G \frac{V_H}{2} \frac{R_A}{R_A + \frac{R_S}{2}}, \quad (1)$$

Where $G$ is the gain of the amplifier stage. The total noise superimposed to the signal reads

$$N \approx 2\left(GN_{\text{in}} + 10^{\frac{NF}{10}}GN_{\text{in}} + \frac{10V_R}{2^8}\right), \quad (2)$$

where the first term represents the amplified sum of the Johnson and pulse generator noises ($N_{\text{in}}$), the second term the noise introduced by the amplifier with noise figure $NF$, and the third term the vertical resolution of the oscilloscope with 8 bits and acquisition range $V_R$ (see Supplementary Note 1 for a detailed derivation of Eqs. 1 and 2). On the basis of Eqs. 1 and 2, we estimate a signal-to-noise ratio $\frac{S}{N} \approx 2.2$ and $\approx 66$ for the single-shot and average traces measured with $V_P = 2.2$ V, respectively. These values are in fair agreement with the actual $\frac{S}{N}$ that characterizes the traces in Figs. 2 and 3. The main contributions to the noise are the $NF$ of the amplifiers (54%) and the resolution of the oscilloscope (30%). The $\frac{S}{N}$ can thus be improved by means of amplifiers with lower $NF$ (1-2 dB, against the 6 dB of our current setup) and oscilloscopes with higher vertical resolution (up to 10-12 bits) or better vertical range.

The temporal resolution is determined by the sampling rate and bandwidth of the oscilloscope as well as by the acquisition mode. In this work, all the traces were acquired in the interpolated real-time mode, which allows for a nominal temporal resolution of $\approx 100$ ps, sufficient to track the dynamics of ns-long pulses. Using an oscilloscope with a higher sampling rate could improve the time resolution down to about 10 ps. The minimal duration of the pulses that can be used to excite the magnetization, on the other hand, is determined by the impedance matching and symmetry of the circuit. In our case, the minimal pulse length is limited to a few ns by the inductive coupling between the wire bonds that connect the sample, which gives rise to over- and under-shoots in the transverse voltage at the rising and falling edges of a pulse (see Supplementary Note 4). This problem can be solved by using optimally-matched rf probes to connect the sample. Ultimately, it is of primary importance that the two branches of the injection (sensing) lines have equal lengths in order to guarantee the synchronization of the injected (sensed) signals. For symmetric branches, the relative delay of the balanced pulses at the center of the Hall cross is determined by the balun divider and is of the order of 1 ps (Ref. [51]). Such a time lag limits the duration of the shortest measurable pulses.

**DISCUSSION**

We have demonstrated a technique to perform time-resolved measurements of the Hall effect and transverse magnetoresistive signals in devices with current flowing in-plane and applied it to investigate with sub-ns resolution the switching dynamics of ferrimagnetic dots induced by SOTs. Our results show that the current-induced magnetization reversal in GdFeCo is characterized by strong stochastic fluctuations of the time required to nucleate a domain. The quiescent phase that precedes the nucleation is a dynamical characteristic that ferrimagnets share with ferromagnets and that has not been reported previously for these materials. The observation of this phase, whose duration and variability are determined by the applied current and in-plane field, implies that the switching process is thermally activated. The corresponding switching delay depends on the combination of two effects. For a given strength of the SOTs and in-plane field, the average duration of the quiescent phase $\bar{t}_0$ is mainly determined by the temperature dependence of the magnetic anisotropy and the rate of increase of the temperature[47]. In this scenario, $t_0$ does not change between switching events and its standard deviation should be of the order of the pulse rise time. In addition to this deterministic process, $t_0$ is influenced by stochastic thermal fluctuations, which cause the spread reported in Fig. 4.

Upon reducing the length of the pulses and increasing their amplitude, the nucleation time can be suppressed to below 1 ns, which results in a minimum of the critical switching energy. Following the initial nucleation phase, the transition between two opposite magnetization states is both fast and monotonic, compatible with the extremely large domain-wall velocity reported for ferrimagnets. However, the reversal is also highly non-deterministic and characterized by a spread of transition times, which deserves further investigation. Overall, our data show that the switching delay time can be rather long in ferrimagnets, unlike the subsequent domain wall motion, which is very fast. The coexistence of these slow and fast phases should be considered in future studies of ferrimagnets to correctly quantify the switching speed.

The sensitivity of the time-resolved Hall measurements is sufficient to perform both average and single-shot measurements, thus providing access to reproducible and stochastic processes. This dual capability combined with the straightforward implementation of our scheme and the widespread availability of Hall experimental probes makes our technique useful for a broad range of studies. The temporal evolution of the transverse voltage can be induced directly by the current, as in this work, or by a different stimulus, like magnetic fields, light or heat, using a pump-probe scheme with a variable delay time between excitation and counter-propagating voltage pulses. In the latter case, the electric current serves uniquely as the probing tool and its duration, amplitude, and waveform can be arbitrarily chosen. As any form of Hall effect or transverse magnetoresistance equally fits our detection scheme, potential applications include time-resolved investigations of electrically- and thermally-generated spin currents and spin torques in magnetic materials, switching of collinear and noncollinear antiferromagnets, as well as time-of-flight detection of skyrmion and domain walls in racetrack devices. Time-resolved Hall effect measurements can also probe the emergence or quenching of symmetry-breaking phase transitions in driven systems. Further, as the Hall response is a quintessential signature of chiral topological states, real-time detection can provide insight into edge transport modes as well as current-induced transitions between quantum Hall and dissipative states.

## METHODS

**Device fabrication.**

The Hall crosses and the dots were fabricated by lithographic and etching techniques. First, the full stack substrate/Ta(3)/Pt(5)/Gd$_{30}$Fe$_{63}$Co$_7$(15)/Ta(3)/Pt(1) (thicknesses in nm) was grown by dc magnetron sputtering on Si/SiN(200) substrate, pre-patterned by e-beam lithography, and subsequently lifted off. A Ti hard mask was defined by a second step of e-beam lithography, electron evaporation, and lift-off. The hard mask protected the circular areas corresponding to the dots during the Ar-ion milling that was used to etch the layers above Pt(5) and define the Hall crosses. Finally, Ti(5)/Au(50) contact pads were fabricated by optical lithography and electron evaporation, followed again by lift-off.

**Electrical setup.**

With reference to Fig. 1, the pulses are produced by a reverse-terminated pulse generator (Kentech RTV40) with variable pulse length (0.3-20 ns, rise time < 0.3 ns) and adjustable polarity, and fed to a directional coupler, which delivers a small portion (-20 dB) of the signal directly to the oscilloscope (trigger). The balanced-unbalanced (balun) power divider (200 kHz – 6 GHz, Marki Microwave BAL-0006) splits the signal into two balanced pulses, with very similar amplitude. Next, the pulses travel to the Hall cross through identical paths. The four bias-Tees next to it combine the rf and dc sub-networks of the circuit, allowing both time-resolved (oscilloscope) and static (lock-in amplifier) measurements. Prior to detection, the transverse Hall potentials are amplified by amplifiers (Tektronik PSPL5865) with 26.5 dB voltage gain, 30 ps rise time and 30 kHz – 12 GHz bandwidth. The oscilloscope is also a Tektronik instrument, with 2.5 GHz bandwidth, 20 GSa/s sampling rate, and 50 Ohm ac-coupled input impedance. A lock-in amplifier (Zurich Instruments MFLI) generates a small low-frequency sinusoidal current ($I_\text{out}$, 100-200 µA, 10 Hz) and demodulates the corresponding static anomalous Hall voltage ($V_\text{in}$). The Hall cross lies on a custom-built printed-circuit board with SMA connections and is contacted electrically by Al wire bonds. The device is located between the pole pieces of an electromagnet, whose magnetic field $B$ can be varied in amplitude and direction within the $xz$ plane.

**Fits of the time-resolved Hall voltage traces.**

We fit the individual normalized switching traces with a piecewise linear function of the form:

$$\text{UP} - \text{DOWN}: y(t) = \begin{cases} 1, & t < t_0 \\ 1 - \dfrac{t - t_0}{\Delta t}, & t_0 < t < t_0 + \Delta t \\ 0, & t > t_0 + \Delta t \end{cases}$$

$$\text{DOWN} - \text{UP}: y(t) = \begin{cases} 0, & t < t_0 \\ \dfrac{t - t_0}{\Delta t}, & t_0 < t < t_0 + \Delta t \\ 1, & t > t_0 + \Delta t \end{cases}$$

for up-down and down-up switching, respectively. We chose a piecewise linear function because of its simplicity and its robustness with respect to the fitting routine as opposed to, e.g., the cumulative function of the Gaussian distribution, which is more prone to errors for small values of $t_0$.

## DATA AVAILABILITY

The datasets generated and/or analysed during the current study are available from the corresponding authors on reasonable request. The data for all of the figures are also available in https://www.research-collection.ethz.ch/, DOI: 10.3929/ethz-b-000458679.

**ACKNOWLDEGEMENTS**

This work was funded by the Swiss National Science Foundation (GrantS No. 200020-172775 and No. PZ00P2-179944), the Swiss Government Excellence Scholarship (ESKAS-Nr. 2018.0056) and the ETH Zurich (Career Seed Grant SEED-14 16-2).

**AUTHOR CONTRIBUTIONS**

P.G., E.G., G.S., and T.D. conceived the experiments. G.S. and V.K. developed the setup and the measurement protocol. C.-H.L. deposited the samples. G.S. fabricated the device, performed the measurements, and analyzed the results. G.S. and P.G. wrote the manuscript. All authors discussed the data and commented on the manuscript.

**COMPETING INTEREST**

The authors declare no competing financial interests.

**ADDITIONAL INFORMATION**

Supplementary information is available in the online version of the paper.

Correspondence and requests for materials should be addressed to G.S. (giacomo.sala@mat.ethz.ch) and P.G. (pietro.gambardella@mat.ethz.ch).

**FIGURES**

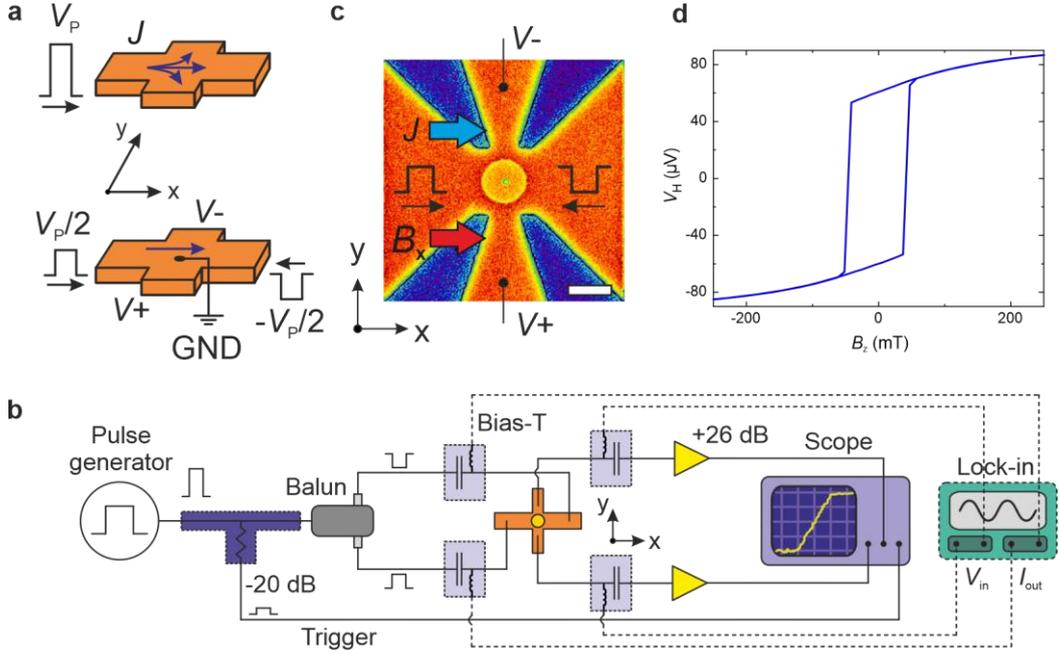

Figure 1. **Experimental setup for time-resolved Hall effect measurements. a,** The injection in a Hall cross of a single pulse with amplitude $V_P$ causes current ($J$) shunting in the transverse sensing line (along $y$ in the upper panel). In contrast, two pulses with opposite polarity ($\frac{V_P}{2}$) that meet at the center of the Hall cross impose a virtual ground, thereby forcing the current to propagate along the main channel (along $x$ in the bottom panel). **b,** Schematics of the rf setup. The initial pulse is fed to a balun divider, which splits the signal into two half pulses with opposite polarity that reach the device at the same instant. The current-induced transverse Hall potentials are amplified and detected by the oscilloscope, triggered by an attenuated portion of the initial pulse. Note that the electric paths traversed by $V_+$ and $V_-$ are symmetric and have equal length in the real setup. The dc sub-network (lock-in amplifier and bias-Ts, dashed lines) allows for the static characterization of the device. **c,** The device is a 1-µm-wide ferrimagnetic GdFeCo dot at the center of a Pt Hall cross, as shown by the false-color scanning electron micrograph. The in-plane magnetic field $B_x$ is collinear to the current. The scale bar corresponds to 1 µm. **d**, Out-of-plane hysteresis loop of a GdFeCo dot measured by the anomalous Hall effect.

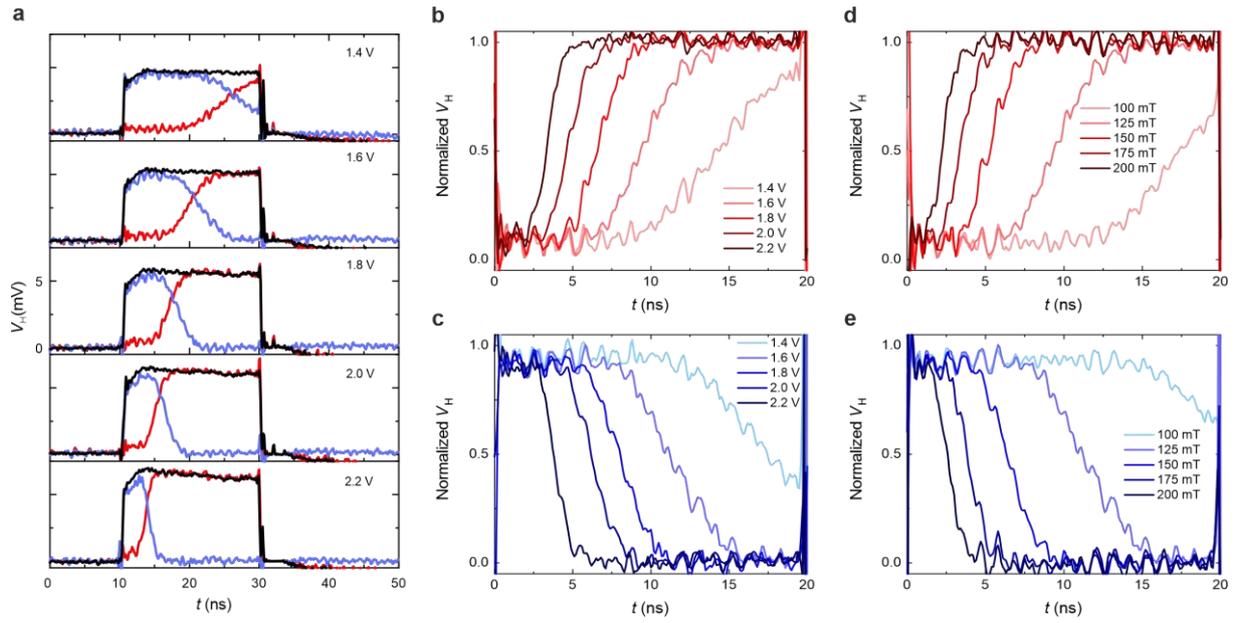

Figure 2. **Switching dynamics of ferrimagnetic dots. a,** Reference (in black) and switching traces of Pt/GdFeCo dots for 20-ns-long voltage pulses of increasing amplitude, showing up-down (blue lines) and down-up (red lines) reversals. The curves are averages of 1000 events. The in-plane magnetic field is 125 mT. **b,c,** Normalized down-up and up-down switching traces at different pulse amplitudes corresponding to the traces in **a**. The current density in the Pt layer corresponding to a pulse amplitude of 1.4 V is ≈ $5.2 \times 10^{11}$ A m$^{-2}$. **d,e,** Normalized down-up and up-down switching traces at different in-plane fields, for pulses with 1.6 V amplitude. In all the measurements the current was positive, whereas the field was positive (negative) in **c,e** (**b,d**).

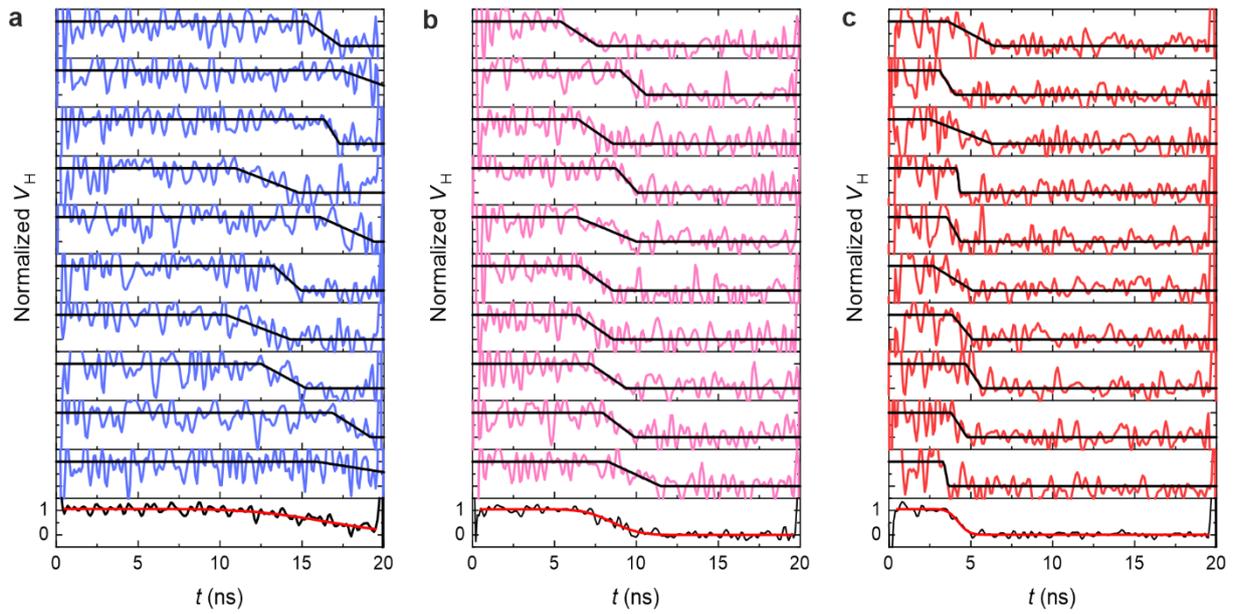

Figure 3. **Single-shot Hall effect measurements.** Normalized single-shot traces of Pt/GdFeCo dots for 20-ns-long pulses. The pulse amplitude is 1.4, 1.8, and 2.2 V in **a, b** and **c**, respectively. The in-plane magnetic field is 125 mT. The pulse amplitude in **a** is close to the threshold switching voltage (see Supplementary Note 3). The black lines are fits to the traces with a piecewise linear function. The bottom-most curve in each graph is the average of the 10 traces above, fitted with the cumulative Gaussian function (red).

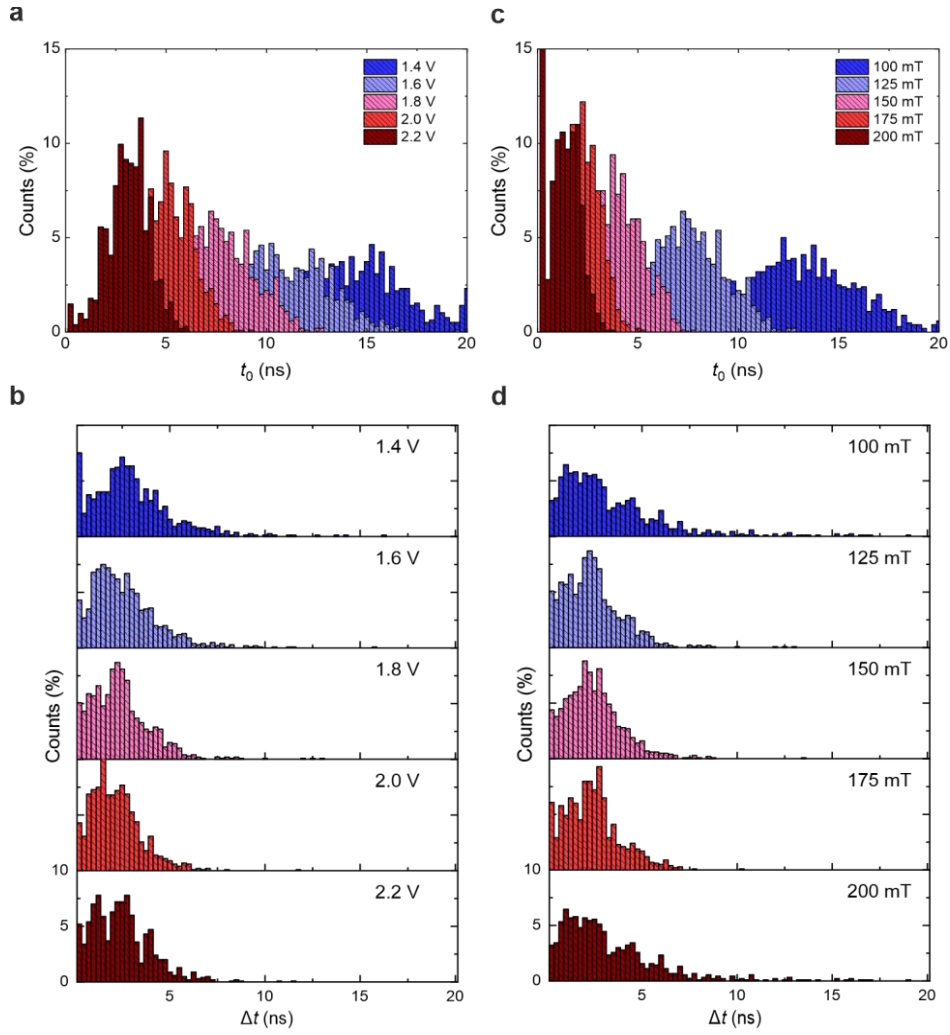

Figure 4. **Distribution of nucleation and transition times. a,b** Percentage distributions of the nucleation time $t_0$ and transition time $\Delta t$ for different amplitudes of 20-ns long voltage pulses, extracted from the fits of the single-shot traces. At 1.4 V, the magnetization does not switch in 22.5% of the events; these events are not included in the plot. The in-plane field is 125 mT. **c,d** Same as **a,b**, for different in-plane fields at a constant pulse amplitude of 1.8 V. At 100 mT, the magnetization does not switch in 9.8% of the events. At 200 mT, the left-most bin includes 24% of the events. This is likely an artifact of the fits due to the limited signal-to-noise ratio of the traces, which causes difficulties in fitting the dynamics close to the rising edge of the pulse. To ease the comparison, in all of the graphs the binning size is 250 ps, larger than the temporal resolution of 100 ps.

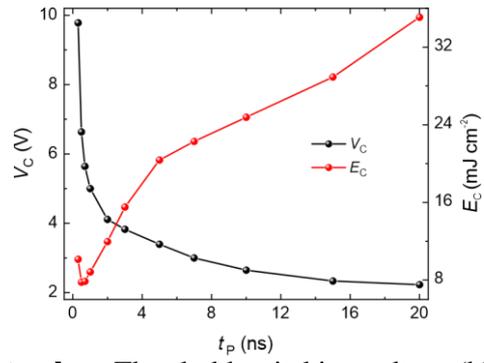

Figure 5. **Switching with short pulses.** Threshold switching voltage (black dots, left scale) and energy density (red dots, right scale) as a function of the pulse length. The critical switching voltage is determined by after-pulse probability measurements as the voltage at which the device switches in 50% of the trials (see Supplementary Notes 3 and 6). The applied in-plane field is 100 mT.

# SUPPLEMENTARY INFORMATION

**Table of contents**

**Supplementary Note 1. Sensitivity of the technique**

The smallest detectable signal is determined by the relative amplitude of the time-resolved anomalous Hall signal and the noise of the electric circuit. In what follows, we estimate the sensitivity of our technique by calculating the anomalous Hall voltage generated by the magnetic dots (see Fig. S1a), the time-resolved amplified voltage, and the superimposed noise.

The anomalous Hall voltage $V_H$ depends on the transverse anomalous Hall resistance $R_{xy}$ and on the current $I_x$, thus it can expressed as

$$V_H = R_{xy} I_x = R_{xy} \frac{V_P}{R_I},$$

where $V_P$ is twice the amplitude $V_P/2$ of the positive (or negative) pulse in Fig. S1b, and $R_I$ the resistance of the injection line. $R_{xy}$ is directly proportional to the anomalous Hall resistivity $\rho_{xy}$, but the comparison with values reported in the literature is not immediate because of geometrical reasons. First, the current distribution is highly inhomogeneous in the Hall cross. Second, most of the current flows through the Pt layer but a small portion enters also the GdFeCo dot and propagates vertically. Third, the anomalous Hall effect does not extend over the entire cross but is limited to the dot area. This geometry is very different from the typical experimental configuration used to measure $\rho_{xy}$, namely a multilayer Hall bar, where the current spreads out in the magnetic layer, which is continuous and extends to the transverse probes, i.e., the sensing line. To account for these differences, we introduce three geometrical parameters. We define the filling factor $F = \pi(\frac{D}{2w})^2$ as the ratio between the areas of the dot and the central portion of the cross (see Fig. S1a), arguing that the Hall signal scales with the magnetic area. In addition, we introduce the sensitivity factor $\varepsilon$, which represents the finite sensitivity of the probes to variations of the electric potential in the cross[1,2]. $\varepsilon$ is determined by the dimension of the Hall cross and by the position of the dot with respect to its center. Finally, we add a parameter $\delta$ that takes into account the non-uniform contribution to the anomalous Hall voltage across the thickness of GdFeCo. The value of $\delta$ is determined by the specific current distribution within the volume of the dot and by the relative weight of volume and interface as sources of the anomalous Hall voltage. Therefore, given the thickness $t$ of the Pt layer, the anomalous Hall voltage reads

$$V_H = \varepsilon \delta F \frac{\rho_{xy}}{t} \frac{V_P}{R_I}. \quad (1)$$

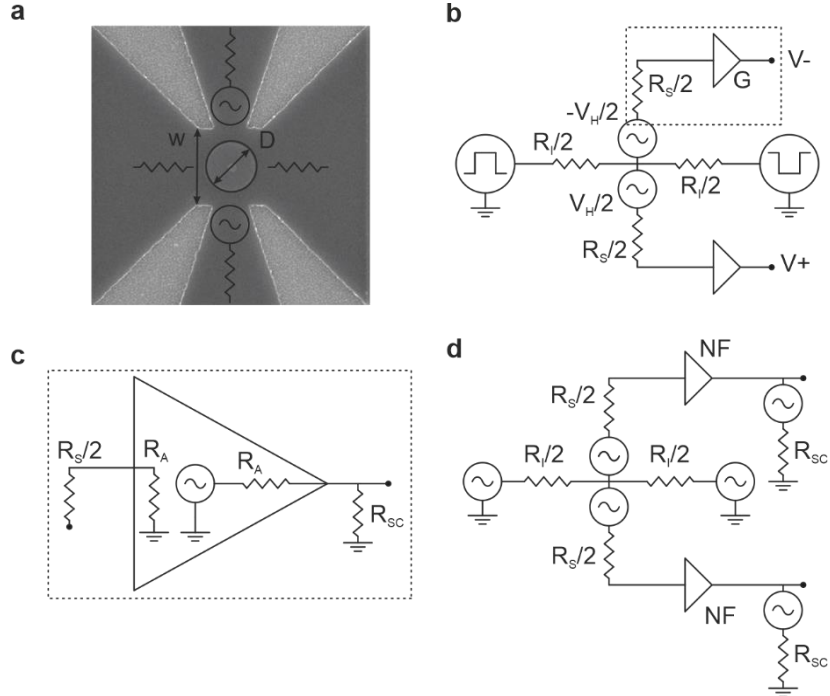

Supplementary Figure S1. **Electrical model of the Hall cross. a,** Scanning electron micrograph of the 1-µm-wide dot and Hall cross, and associated resistors. $w$ and $D$ are the width of the Hall cross and the diameter of the dot, respectively. **b,** Equivalent electric circuit of the Hall cross, with the associated resistors, voltage sources, and amplifier. The dashed rectangle corresponds to the schematic in **c**, which represents the equivalent model of the amplifier. **d,** Model of the Johnson noise: every resistor is replaced by an ideal noise-free resistor and a noise voltage source.

This voltage is then amplified and measured in real time. Figure S1b presents the equivalent electric circuit of the Hall cross. We model the device with four resistors ($2 \times \frac{R_I}{2}, 2 \times \frac{R_S}{2}$) connected through a central node. The resistors represent the two branches of the injection line with total resistance $R_I$, along which the pulses are injected, and the two branches of the sensing line with resistance $R_S$ used to probe the anomalous Hall voltage. Since the resistance difference between the two branches of the sensing (injection) line is a few Ohm at most, we assume for simplicity that the branches are equal in pairs. For the sensing line, this hypothesis is equivalent to considering an ideal offset-free transverse voltage. The anomalous Hall effect can be modelled by two voltage supplies of opposite sign ($\pm \frac{V_H}{2}$) placed along the two sensing branches. At the centre of the cross, the counter-propagating pulses enforce a virtual ground. Then, the differential Hall signal $S$ measured at the input ports of the oscilloscope is the result of the amplified voltage partition between $\frac{R_S}{2}$ and the input resistance of the amplifier $R_A$:

$$S = 2G \frac{V_H}{2} \frac{R_A}{R_A + \frac{R_S}{2}} \quad (2)$$

Here, the amplifier is treated as a simple ideal amplifying stage with gain $G$ and $R_A = 50$ Ohm input and output impedances (see Fig. S1c). The input resistance of the oscilloscope is also $R_S = 50$ Ohm.

The measured amplified signal is accompanied by noise, which originates mainly from the Johnson noise of the resistors ($N_J$), the noise of the pulse generator ($N_P$) caused by its output impedance, the resolution of the oscilloscope $N_{SC}$, and, above all, the noise figure ($NF$) of the amplifiers. Additional noise sources are the passive electric devices present in the circuit (bias-Tees, couplers, balun divider). Moreover, the wire bonds and our printed circuit board pick up electromagnetic disturbances from the environment. However, these extra noise contributions are negligible compared to $N_J$, $N_P$, $N_{SC}$, and $NF$. We model the Johnson noise by replacing each resistor in Fig. S1b with the equivalent Thevenin circuit, comprising of an ideal resistor of the same resistance $R$ and voltage source $N_J = \sqrt{2} N_{rms} = \sqrt{8k_B T \Delta f R}$, with $k_B T \approx 4.1 \times 10^{-21}$ J the thermal energy and $\Delta f \approx 50$ MHz the bandwidth (20 ns pulses). The resulting equivalent noisy circuit is sketched in Fig. S1d. It can be simplified by condensing the contributions of all the resistors into a single effective resistance $R_{eff}$:

$$R_{eff} = \left( \frac{\frac{2R_A R_I}{2R_A + R_S + 2R_I}}{\frac{R_I}{2} + \frac{(R_A + \frac{R_S}{2})R_I}{2R_A + R_S + 2R_I}} \right)^2 \frac{R_I}{2} + \left( \frac{R_A + \frac{2R_A R_I}{8R_A + 4R_S + 2R_I}}{R_A + \frac{R_S}{2} + \frac{(R_A + \frac{R_S}{2})R_I}{4R_A + 2R_S + R_I}} \right)^2 \frac{R_S}{2}.$$

Then, the input noise to each amplifier is

$$N_{in} = \sqrt{8k_B T \Delta f R_{eff}} + \frac{R_A}{R_A + \frac{R_S}{2}} N_P.$$

Considering also the Johnson noise of the oscilloscope's input impedance $R_{SC}$ and the digital-to-analogue quantization, the total noise superimposed to the signal reads

$$N = 2\left(GN_{in} + 10^{\frac{NF}{10}}GN_{in} + \sqrt{8k_BT\Delta fR_{SC}} + \frac{10V_R}{2^8}\right), \quad (3)$$

where the first term represents the amplified sum of the Johnson and pulse generator noises, the second term the noise introduced by the amplifier, and the third term the Johnson noise of the input impedance $R_{SC}$ of the oscilloscope. Strictly speaking, the last contribution in Eq. (3) is not noise, but the intrinsic finite sensitivity of the oscilloscope. This resolution is determined by the number of bits (8) and divisions (10), and the voltage range ($V_R$). Finally, the factor 2 is due to the mathematical subtraction of the amplified $V_+$ and $V_-$.

Equations (1)-(3) can be used to estimate the signal-to-noise ratio ($S/N$) and the sensitivity of the setup. In our case: $D$ = 1000 nm, $w$ = 1500 nm, $R_I$ = 360 Ohm, $R_S$ = 806 Ohm, $R_{SC}$ = 50 Ohm, $R_A$ = 50 Ohm, $G$ = 20 (26 dB), $NF$ = 6 dB, $N_P$ = 9 µV, $V_R$ = 7 mV, $\varepsilon$ = 0.4 (Ref. [2]). These values lead to: $R_{eff}$ = 16 Ohm, $N_{in}$ = 6.1 µV, $F$ = 0.35. We assume a pulse with amplitude $V_P$ = 2.2 V, an anomalous Hall resistivity $\rho_{xy}$ = 10 µOhm cm (Refs. [3,4]), and $\delta$ = 0.21, the latter being chosen to match the experimental $R_{xy}$ = 0.6 Ohm. Thus, we obtain $V_H$ = 3.6 mV and $S$ = 7.9 mV. This last result is in good agreement with the experimentally measured value (cf. Fig. 2a). The total noise amounts to $N$ = 1.8 mV. This value is to a large extent (up to 54%) determined by the noise figure of the amplifier, which intensifies the Johnson noise $N_{in}$ of the circuits. This noise is expected to become more severe as the pulse length is reduced, i.e., the bandwidth is enlarged. For a 1-ns long pulse, it may increase by 4-5 times. The second largest contribution (30%) originates from the signal quantization, whereas the contribution of $R_{SC}$ is negligible. On the basis of these figures, we estimate that the signal-to-noise ratio of a single measurement is of the order of $S/N \approx 4.4$. Since the time traces are obtained by subtraction of two measurements (see Supplementary Note 4), the $S/N$ of the individual time trace (single-shot measurements) reduces to $\approx 2.2$. By averaging over 1000 switching traces, the ratio can be improved by a factor of 30, which gives $S/N \approx 66$. This estimate matches reasonably well the actual signal-to-noise ratio of the average traces in Fig. 2a (bottom panel, 2.2 V pulse amplitude), which have about 6.5 mV and 0.15 mV signal and root-mean-square noise amplitude, respectively.

These considerations explain why our technique is advantageous. Without the compensation of the pulses at the centre of the Hall cross, the transverse signals $V_+$ and $V_-$ are of the same order of magnitude as the injected pulse, e.g., 1 V. The magnetic signal is thus a tiny variation on the order of a few mV on top of the large background. In such conditions, a much higher range $V_R$ is required to accommodate the entire signal into the available divisions of the oscilloscope. As a consequence, the finite vertical resolution becomes dominant over the rest of the noise and masks the magnetic signal. Sourcing the oscilloscope with the differential signal $V_+ - V_-$ would definitely improve the resolution by removing part of the background. Still, this approach would not solve completely the problem, because of the asymmetries between the sensing branches. In contrast, our technique minimizes the current spread and hence allows for exploiting the full acquisition range of the oscilloscope to probe only the magnetic signal.

This analysis suggests also a few directions for further improvements. In the first place, the device geometry and the materials (thickness, resistivity) could be designed to maximize $V_H$. For example, the anomalous Hall resistance could be enhanced by increasing the ratio between the width of the sensing arms and the dot diameter[5,6], so as to increase the factor $\varepsilon$. Likewise, the central area of the cross should be made the smallest possible, compatibly with the dot size. This optimization becomes fundamental when downscaling the devices to sub-µm dimensions. However, the device optimization is not free from constraints because the anomalous Hall voltage, the current density required to induce the magnetization switching, the geometry of the Hall cross, and its resistance are not independent. For instance, the device miniaturization, which would enlarge $V_H$, would also increase the resistance of both the injection and sensing lines, hence the Johnson noise. Therefore, an alternative option is the optimization of the setup. At the present stage, the critical source of noise in our circuitry is the voltage amplifier. With all other parameters fixed, amplifiers with a 1 dB noise figure are expected to provide $S/N = 3.5$ for the single-shot traces. Additionally, the subtraction of $V_+$ and $V_-$ prior to detection by the oscilloscope should improve the $S/N$ by permitting the reduction of $V_R$. If $V_R$ is reduced to the minimum of our oscilloscope (2 mV), then the $S/N$ would further increase to 5.4. The subtraction could be done with an additional balun used in the opposite configuration, namely, with the input signals $V_+$ and $V_-$ connected to the inverting and non-inverting ports of the device.

**Supplementary Note 2. Temporal resolution of the technique**

As described in the main text, the temporal resolution is determined by the sampling and by the acquisition mode (real time, interpolated real time, etc.). In this work, the traces were acquired in the interpolated real-time mode, which allows for a nominal temporal resolution of $\approx$ 100 ps, sufficient to track the dynamics of ns-long pulses. For shorter pulses, the nominal resolution could be improved to a few ps by using a faster oscilloscope. We note that the other elements of the circuit and the cabling may distort the shape of the electrical excitation if their transfer function does not match the required frequency range, but they do not influence the temporal resolution. Instead, it is of primary importance to ensure the equal length and symmetry of the injection (sensing) lines of the circuits to guarantee the synchronization of the injected (sensed) signals.

The shortest traces that we could reliably measure correspond to 2-3 ns-long pulses. This limitation has a different "extrinsic" origin than the circuit components, namely the geometry of the Hall cross, which was not specifically designed for transmitting rf pulses, and, above all, the use of wire bonds to contact the device, which are inductively coupled. As a consequence, the raw traces have edge spikes with about 1 ns FWHM (see Fig. S4c) that complicate the analysis of the magnetic traces for pulses shorter than 1 ns. The replacement of the wire bonds with rf probes would improve the transmission of sub-ns pulses. We stress that these limitations affect the length of the pulses, but not the temporal resolution, which remains 100 ps and can be independently improved.

**Supplementary Note 3. Sample characterization**

Figure S2a reports the hysteresis loops of a $Gd_{30}Fe_{63}Co_7$ device as probed by static measurements of the anomalous Hall resistance, with field applied perpendicular to the plane (polar angle = 0°) and almost in plane (89°). The sense of rotation of the hysteresis loop indicates that the magnetization is dominated by the transition metals Fe and Co. The GdFeCo layer has perpendicular magnetic anisotropy, with an effective anisotropy field of the order of 300 mT. The saturation magnetization was estimated to be 25 kA/m using SQUID magnetometry performed on a full film sample. The device can be reliably switched between the up and down states by bipolar electric pulses in presence of an in-plane magnetic field collinear with the current direction, as typical of spin-orbit torques (see Fig. S2b).

We note that the GdFeCo devices studied here belong to a batch of samples with variable Gd concentration that cross the magnetization compensation temperature. However, we found that the fabrication steps alter the properties of the devices with respect to those of the full films. This undesired change is one of the limitations of amorphous ferrimagnets, which are particularly sensitive to standard operations such as the ion milling and the resist baking. These issues have already been observed by other groups (see e.g., Ref. [7–9]) and are possibly caused by the selective oxidation or migration of the rare-earth atoms[10]. Our estimate, based on the variation of the magnetization compensation temperature with the Gd concentration (about 30 K every 1%), is that the magnetization compensation temperature is around 250 K. Because of Joule heating during pulsing, we are confident that the magnetization of our devices is always "FeCo-like" for the time-resolved Hall effect experiments reported in this work, which were all performed in ambient conditions.

In order to determine the working point required to induce the switching, we measured the probability of switching as a function of in-plane magnetic field and pulse amplitude. To this aim, we used the dc sub-network of the circuit shown in Fig. 1 in the main text. The procedure was the following. We applied a sequence of set-reset rf pulses with identical length and amplitude but opposite polarity. The variation of the transverse dc resistance before and after each pulse was compared with the anomalous Hall resistance to assess the outcome of the pulse: if the variation was larger than 75 % of this reference, we considered that the pulse succeeded in switching the magnetization. Every pulse sequence comprised 50 set-reset pairs of pulses, and the switching probability was defined by the ratio of successful pulses to 50. We repeated this procedure for different pulse lengths, amplitudes, and fields, as reported in Fig. S3a-d. As expected, the minimum voltage for 100% switching decreases as the field or the pulse length are increased

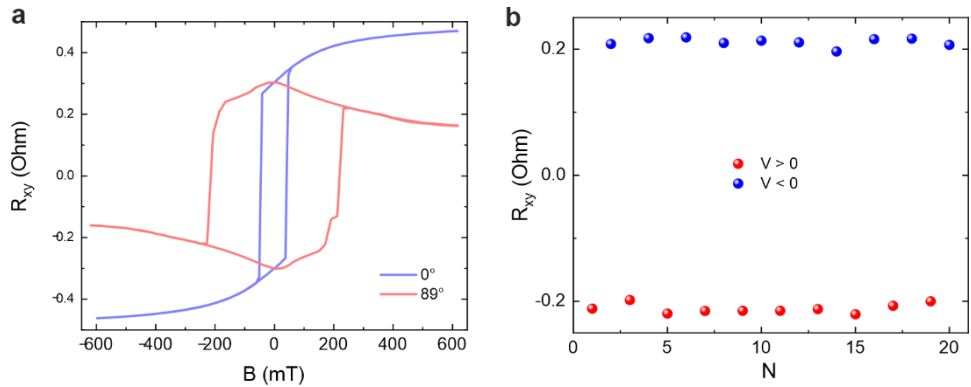

Supplementary Figure S2. **Sample characterization. a,** Hysteresis loops measured by the anomalous Hall resistance with the field applied out of plane (0°) and in plane (89°). **b,** Switching of the magnetization by a sequence of positive set (V > 0) and negative reset (V < 0) pulses with length of 20 ns and amplitude of 1.6 V. The in-plane field was 150 mT. Note that the switching amplitude is smaller than the anomalous Hall amplitude in **a** because of the tilt induced by the applied field (cf. with the red trace in **a** at 150 mT).

**Supplementary Note 4. Measurement protocol and analysis of raw signals**

In an ideal scenario, the signal measured by the oscilloscope should approximately resemble a "rectangle", that is, it should replicate the temporal profile of the applied electric pulse. In such a case, the amplitude of the signal (height of the rectangle) would already represent the measurement of the magnetization state. If the magnetization was in equilibrium, the amplitude would remain constant, to a high or low level in dependence of the up or down orientation of the magnetization. During the switching, instead, the trace would transition from one level to the other. However, spurious nonmagnetic contributions alter the ideally-rectangular profile of the measured signal. These contributions have multiple origins. First, the edge of the pulses have large spikes caused by the inductive coupling between the wire bonds and the electric contacts of the PCB. Second, the device itself, which is not adapted to radio frequencies, distorts the pulses and hence the measured signal. In addition, the voltage amplifier introduces high-frequency oscillations. Finally, as shown in Fig. S4a,b, the voltage difference between the inverted (I) and non-inverted (NI) pulses at the output of the balun divider is several mV, which is about 1% of the pulse amplitude. This component is quite small with respect to the input pulse. Yet, the unbalance causes a residual current leakage through the transverse arms which adds a small voltage offset (comparable to or smaller than the magnetic signal). Therefore, the magnetic signal is better extracted from the raw traces by comparing measurements of a reference and the switching and removing the non-magnetic part. In fact, every trace of the same type as in Fig. 2b-e of the main text results from the combination of two measurements. The procedure that we adopt to isolate the magnetic signal is the following[11].

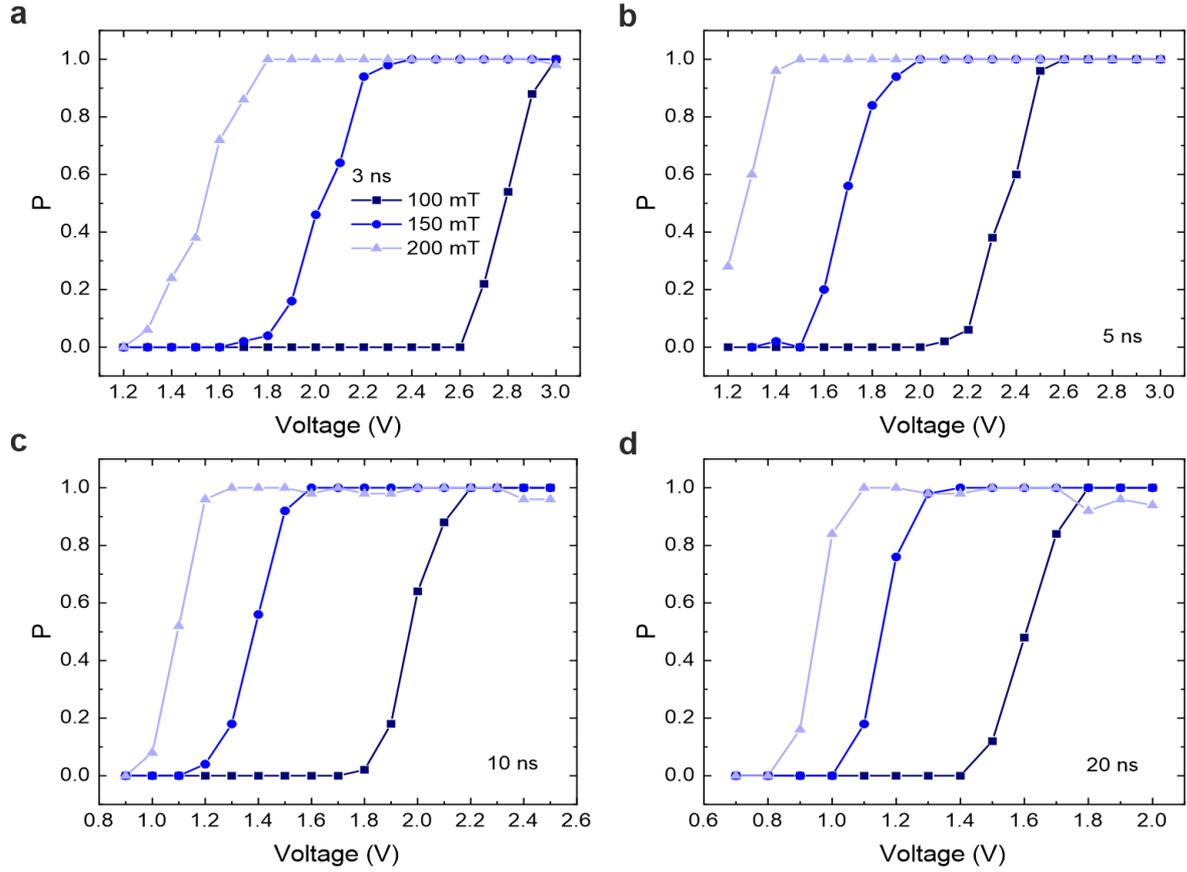

Supplementary Figure S3. **Switching probability. a-d,** After-pulse switching probability as a function of pulse amplitude, for different pulse lengths and in-plane fields.

First, in the presence of a positive in-plane magnetic field, we acquire a background signal by repeatedly injecting identical pulses with the same current direction and amplitude. In these conditions, the magnetization remains in the equilibrium state, which is determined by the field direction and the sign of the spin-orbit torques defined by the current polarity. The latter equals the polarity of the pulse travelling along $+x$, that is, from left to right in Fig. 1a in the main text. The differential voltage $S$ measured during each pulse is nominally always the same, but we average over multiple pulses, typically 5000, to reduce the noise. Then, we repeat the same step for the opposite field direction and the same current polarity, to acquire the background signal corresponding to the opposite equilibrium state (see Fig. S4c). By definition, all the undesired contributions do not change with the magnetic configuration of the device, hence they can be removed by subtracting the two signals. Their difference yields the net magnetic contrast: reference trace = Background ($B < 0$) – Background ($B > 0$). This is the black trace shown in Fig. S4d as well as in Fig. 2a. Since for $V > 0$ and $B < 0$ ($B > 0$), the magnetization remains in the up (down) state, corresponding to positive (negative) anomalous Hall voltage, the reference trace so defined has positive sign.

Next, we acquire the signal corresponding to the switching of the magnetization by slightly varying the procedure, that is, by delivering a train of set-reset pulses with alternating polarity. Now, at each pulse the current direction changes and so does the magnetization. For example, for positive field, the positive current causes the up-down switching, whereas the successive negative current induces the down-up reversal. By averaging over 1000 pulses of the same polarity, we acquire the green signal shown in Fig. S4c (a positive in-plane field is applied). It coincides initially with the signal for Background ($B < 0$) (magnetization up) and during the pulse it transitions to the signal for Background ($B > 0$) (magnetization down). Therefore, similarly to the reference trace, the signal $S$ associated to a switching event is combined with one of the two backgrounds: switching trace = S ($B > 0$) – Background ($B > 0$). The application of this procedure leads to the blue trace in Fig. S4d as well as to the traces in Fig. 2. The ≈ 0 mV (≈ 5 mV) trace level identifies the uniformly-magnetized down (up) state, whereas any deviation of the traces from the top and bottom levels correspond to a tilt of the magnetic moments or to a multi-domain configuration. Finally, the normalization of the switching trace to the reference trace provides the purely-magnetic time traces (cf. Fig. 2b-e in the main text). The same identical approach is used for detecting single-shot events, with the only difference that, instead of averaging, every single switching signal is recorded. The procedure that we adopt to measure and remove the background signal is very similar to that reported in Ref. [12]. Therefore, our measurement protocol is comparable to that of standard time-resolved Hall measurements.

Finally, we note that the reference trace can be acquired by using protocols different from ours, which is adapted to the specific case of spin-orbit torque switching. For example, the background signals of perpendicularly-magnetized samples could also be acquired by fixing the magnetization with out-of-plane fields. If the polarity of the current has an effect, a reference could also be obtained by comparing background signals measured with opposite current polarity. Alternatively, the signal measured with a low-amplitude pulse could be used as background: under the assumption that the low amplitude does not produce magnetic changes, the corresponding trace could be subtracted from a higher-amplitude trace after proper rescaling. In antiferromagnets, repeated pulses produce a memristive-like switching. Then, the background trace could be obtained after applying a sequence of repeated pulses that saturate the read-out signal to the maximum (or minimum) level. Therefore, in general, the measurement protocol can be adapted to the specific application.

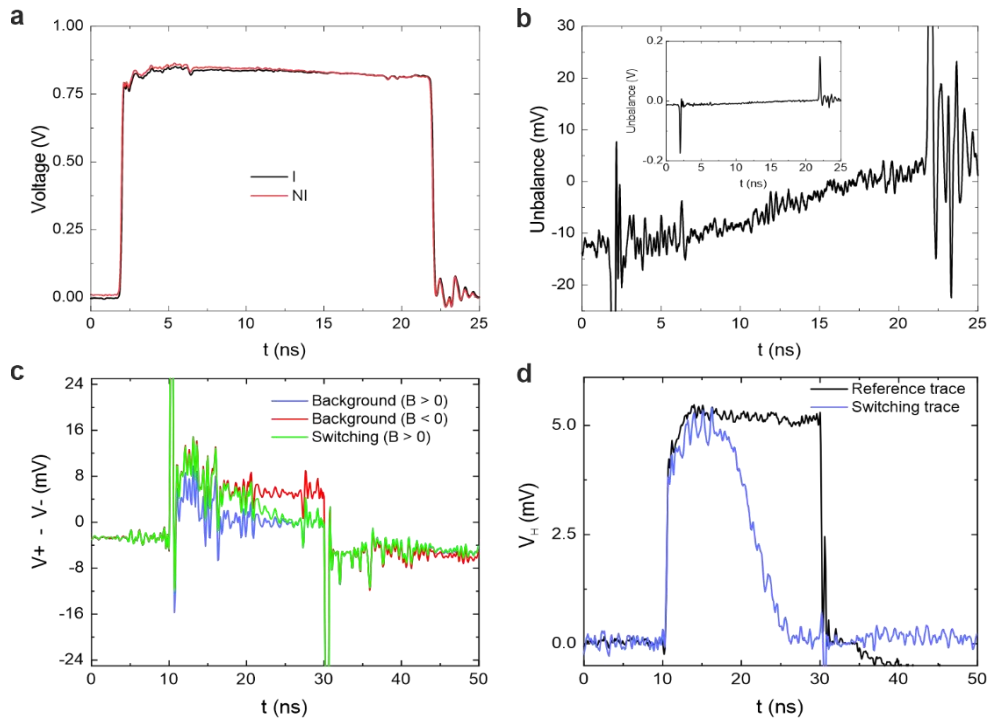

Supplementary Figure S4. **Analysis of raw signals. a,** Inverted (I) and non-inverted (NI) pulses at the outputs of the balun divider, for a 20 ns, 1.6 V input pulse; the sign of the I pulse has been inverted for comparison. **b,** Close-up of the difference between the I and NI pulses (I - NI). Inset: full voltage difference between the two pulses. **c,** Average raw electric signals corresponding to the background, for the two in-plane field directions, and to the switching (for positive field). **d,** Reference and switching traces obtained by subtraction of the signals in **c**.

**Supplementary Note 5. Compensation of resistance offsets.**

Our technique does not imply a more complex circuit or measurement protocol than traditional differential Hall measurements. For comparison, we consider the work by Yoshimura et al. (Ref. [12]). In our setup, we included DC components to simultaneously access the static electric and magnetic properties of the devices. Once the DC subnetwork, which is not necessary for time-resolved measurements, is removed, the sole difference between the differential Hall measurement presented in Ref. [12] and our technique is the balun divider. The balun is a simple, small, and affordable component that does not require any power supply and easily fits into any electrical setup.

As an additional advantage, our technique allows for compensating possible resistive offsets that are caused by the imperfect fabrication or are intrinsic to asymmetric devices. To prove this point, we have measured the raw electrical signals corresponding to the "up" and "down" magnetization states in a Hall bar device with two off-centered Hall crosses (see Fig. S5a). In contrast to the symmetric Hall cross considered in the manuscript, in this device the electric potentials determined by the two pulses at the center of the right Hall cross are different because of the asymmetric resistance load. As a result, the current does flow in the transverse arms and the signals measured on the oscilloscope present a finite offset (see Fig. S5b). Since this offset is not negligible, to acquire the signals we could not use the the maximum vertical resolution of the oscilloscope. Such problems can be circumvented by correcting the pulses amplitudes to enforce the virtual ground at the position of the Hall cross. In the specific case discussed here, we added a 4 dB attenuator along the direction of the negative pulse. Thanks to this adjustment, the vertical offset was removed from the raw signal, which allowed us to exploit the highest vertical resolution of the oscilloscope. Therefore, our technique does not require the device under test to be longitudinally symmetric. Although we do not have at our disposal devices with asymmetric transverse Hall arms, we believe that transverse resistance offsets could be compensated in the same way as for the longitudinal offset. Since commercial attenuators provide attenuation steps as small as 0.5 dB (= 0.944), the amplitude of the pulses can be tuned with rather large precision. This capability is a specificity of our technique, for no such countermeasures can be taken in standard differential Hall measurements.

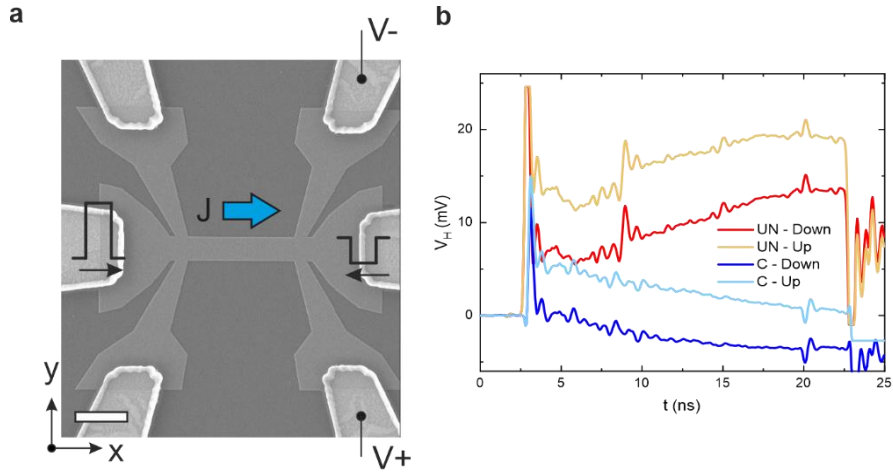

Supplementary Figure S5. **Compensation of resistance offsets. a)** Hall bar with off-centered Hall crosses. The anomalous Hall effect is measured in the right Hall cross. The negative pulse moving from right to left is attenuated by 4 dB compared to the positive pulse. The scale bar corresponds to 4 µm. **b)** Raw differential Hall voltage $V_H = V^+ - V^-$, with uncompensated (UN) and compensated (C) resistance offset, corresponding to the up and down magnetization states for current pulses that do not induce switching.

**Supplementary Note 6. Switching with short pulses.**

The measurements presented in the main text were performed with 20-ns-long pulses. These relatively long pulses allow us to clearly identify the different phases of the dynamics. In Fig. S6 we present additional average time-resolved measurements performed with 5-ns-long pulses. At the largest pulse amplitude the nucleation time is reduced down to about 800 ps. This decrease is consistent with the after-pulse probability measurements shown in Fig. S7a, which shows the switching probability measured as a function of the pulse amplitude and length for a constant in-plane field of 100 mT. The plot demonstrates that deterministic switching can be obtained with pulses as short as 300 ps, which implies quenching of the nucleation time at sufficiently high pulse amplitudes. From Fig. S7a we extracted the threshold switching voltage, defined as the voltage at which the device switches in 50% of the trials, and plotted it against $1/t_P$ in Fig. S7b (see also Fig. 5 in the main text). Below approximately 5 ns, the voltage increases linearly with the inverse of $t_P$, which is a signature of the intrinsic regime where the switching speed depends on the rate of angular momentum transfer from the current to the magnetic layer. On the other hand, the different dependence for $t_P > 5$ ns reveals the importance of thermal effects for the typical pulse lengths used in this study ($t_P = 20$ ns).

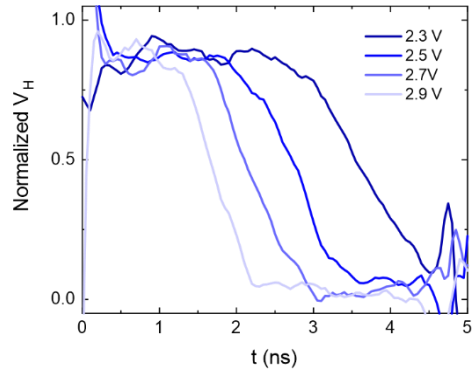

Supplementary Figure S6. **Switching with 5-ns pulses.** Normalized average traces showing the up-down magnetization switching with 5 ns-long pulses of different amplitude. Both the current and the in-plane 125 mT field were positive.

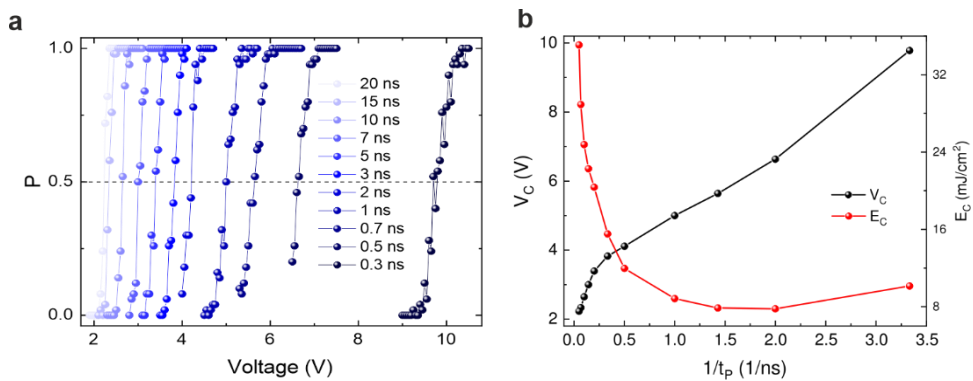

Supplementary Figure S7. **Switching as a function of pulse length. a)** Dependence of the switching probability on the pulse amplitude for different pulse lengths. Each point is the result of 50 trials. The applied in-plane field was 100 mT. Note that these measurements were performed on a different device than that used for the time-resolved measurements but they were fabricated at the same time from the same layer. **b**, Threshold switching voltage (black dots, left scale) and energy density (red dots, right scale) as a function of the inverse pulse length. The critical switching voltage is determined from **a** as the voltage at which the device switches in 50% of the trials.

**Supplementary References**